\documentclass[journal]{IEEEtran}
\usepackage{graphicx}
\usepackage{diagbox}
\usepackage{makecell}
\usepackage{bm}
\usepackage{siunitx}
\usepackage{url}

\begin{document}
%
\title{Performance investigations of two channel readout configurations on the cross-strip cadmium zinc telluride detector
}

\author{Yuli Wang,~\IEEEmembership{Student Member,~IEEE} 
\thanks{*This draft is not under peer-review and was conducted in UCSC.
}
\thanks{*Y. Wang are with the department of biomedical engineering, Johns Hopkins University, Baltimore, 21204.} 
}

\maketitle

\pagenumbering{gobble}

\begin{abstract}
Multiple application-specific integrated circuits (ASIC) are required for the detectors if their readout channels are larger than that of ASIC channels. For a system with such a readout scheme, there is a need to configure channels among ASICs to achieve the lowest electronics noise and highest count rate. In this work, experiments were performed to investigate the performance of two different readout configurations between two ASICs in a cross-strip cadmium zinc telluride detector. A lower electronic noise level, better FWHM energy resolution performance, and higher count rate was found for the anode electrode strips with each ASIC allocating half of the detector area, when compared to allocating each ASIC channel to alternate anode channels. The average electronics noise levels were reduced to $12.61 \pm 0.48\,keV$ units (anode) and $26.16 \pm 3.03\,keV$ units (cathode) for the half-half configuration. The energy resolution of the half-half configuration is $1.65\% \pm 0.05\%$ compared to that of the alternate configuration around $1.90\% \pm 0.06\%$. Charge sharing and scattering play a role in the different count rates, and the count rate of the half-half configuration is $43.9\%$ higher than that of the alternate configuration.
\end{abstract}

\section{Introduction}

\IEEEPARstart{C}{adimum} Zinc Telluride (CZT) detector has gained considerable popularity for applying in positron emission tomography (PET) systems to detect the ionizing radiation \cite{peng2010design,gu2011study,yin2012study,abbaszadeh2016characterization,9534780}. The attention towards CZT is attracted mainly due to their properties of good energy resolution, high inherent spatial resolution, the possibility to easily achieve high packing fraction ($\sim $ 99\%), and the direct detection of gamma photon \cite{zhang2005feasibility,gu2011study,zhang2012characterization,groll2016hybrid}. So far, by stacking a few CZT detectors together, $\sim $ 1 mm spatial resolution of CZT-based small animals or organ-dedicated PET systems have been demonstrated \cite{gu2011study,abbaszadeh2016characterization}. However, the performance of a large volume of CZT detectors-based PET systems are still not well studied, especially the quantitative analysis of the crosstalk issue among a large amount of CZT detectors, the extra electronics noise introduced by flexible circuit and its corresponding bonding issue.

Currently, our lab is developing an ultra-high-resolution dedicated head-and-neck PET system using a large volume of CZT detectors \cite{wang2021back,zhang2020penalized,wang2021high,wang2021electronic,wang2021further,wang2019two}. A two-panel geometry system design is proposed, which aims to improve the detection sensitivity and make the patients feel more comfortable during the scan. Each panel is with 20$\times$15 cm \textsuperscript{2} geometry dimension, which is formed by stacking 150 pieces of 40 mm × 40 mm × 5 mm monolithic CZT detectors together in 5 columns $\times$ 30 rows with “edge-on" detector arrangement. The “edge-on" arrangement could significantly improve each panel's packing fraction to 99\% and 40 mm thickness of the CZT detector could results in a greater 86\% intrinsic detection efficiency for 511 keV photon \cite{levin2006impact,habte2007effects}.

Besides, compared to the scintillation detector, the spatial resolution of the CZT detector is not limited by the manufacturing ability of minuscule crystal elements, but is directly determined by the size of deposited electrode patterns. After embracing this advantage of CZT, we developed our CZT detector with cross-strip electrode patterns \cite{lee2010development,gu2014study}, consisting of 39 anode strips with 100 $\mu$m width, 38 steering strips with 100 $\mu$m width and 8 cathode strips with 4900 $\mu$m width. The cross-strips design could further help us dramatically reduce the number of readout channels \cite{gu2014study} when compared to the pixilated electrode CZT detector with a similar spatial resolution (2n versus n\textsuperscript{2}). The less required number of readout channels could bring benefits in terms of data acquisition bandwidth and electronics thermal management.

A large number of research has focused on optimizing the coincidence timing performance of CZT detector for PET \cite{meng2005exploring}, investigating the induction, propagation, and collection of charge within CZT \cite{komarov2009simulation,kim2011charge,komarov2012investigation} and developing efficient readout electronics with large channel numbers for CZT \cite{abbaszadeh2016characterization,yue2017multi,wang2021back}. In this paper, we will focus on investigating the performance of the two different readout configurations in cross-strips CZT.

Overall, in this paper, we are going to investigate the aforementioned three challenges to improve the detection performance of our large-volume CZT-based PET system and aim to provide guidance for future works regarding building a large-volume CZT system.

\begin{figure}[ht]
\begin{center}
\includegraphics[scale=0.45]{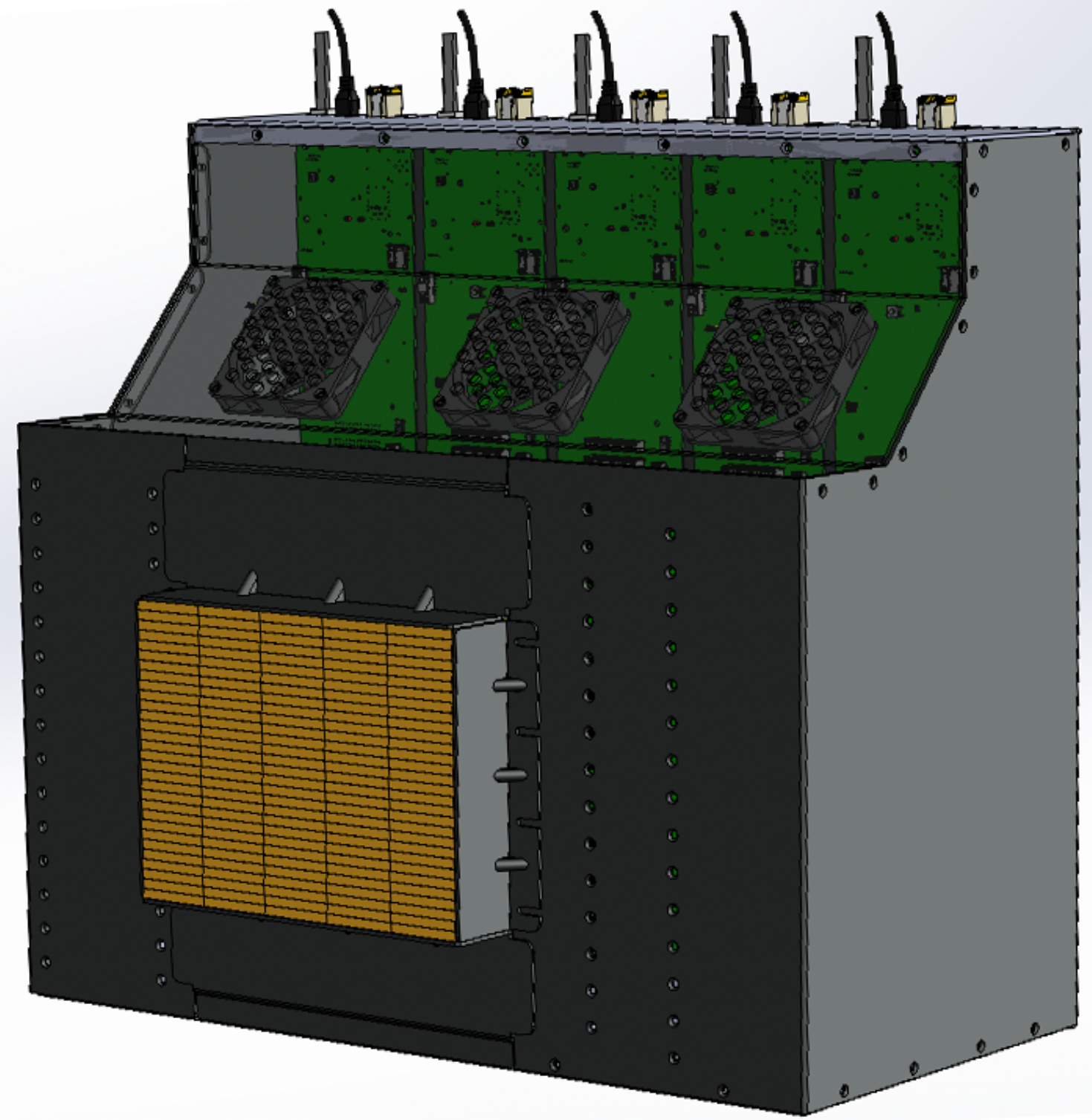}
\end{center}
\caption{Schematic of one-panel system with 30-row by 5-column of 4 cm $\times$4 cm CZT detectors. Each panel incorporates the CZT detectors assembled with flexible circuits, the front-end electronics, and the back-end electronics. The two-panel system contains two adjustable panels, an axial head holder, and other mechanical supports.}
\label{system_schematic}
\end{figure}

\section{Materials and Methods}
\label{sec:Materials_Methods}

This section summarizes the design of the CZT detector module and electronic readout system. Then the system follows the presented methods for the system-level characterization and investigation.

\subsection{Cross-strips CZT detector}

\begin{figure}[ht]
\begin{center}
\includegraphics[width=\columnwidth]{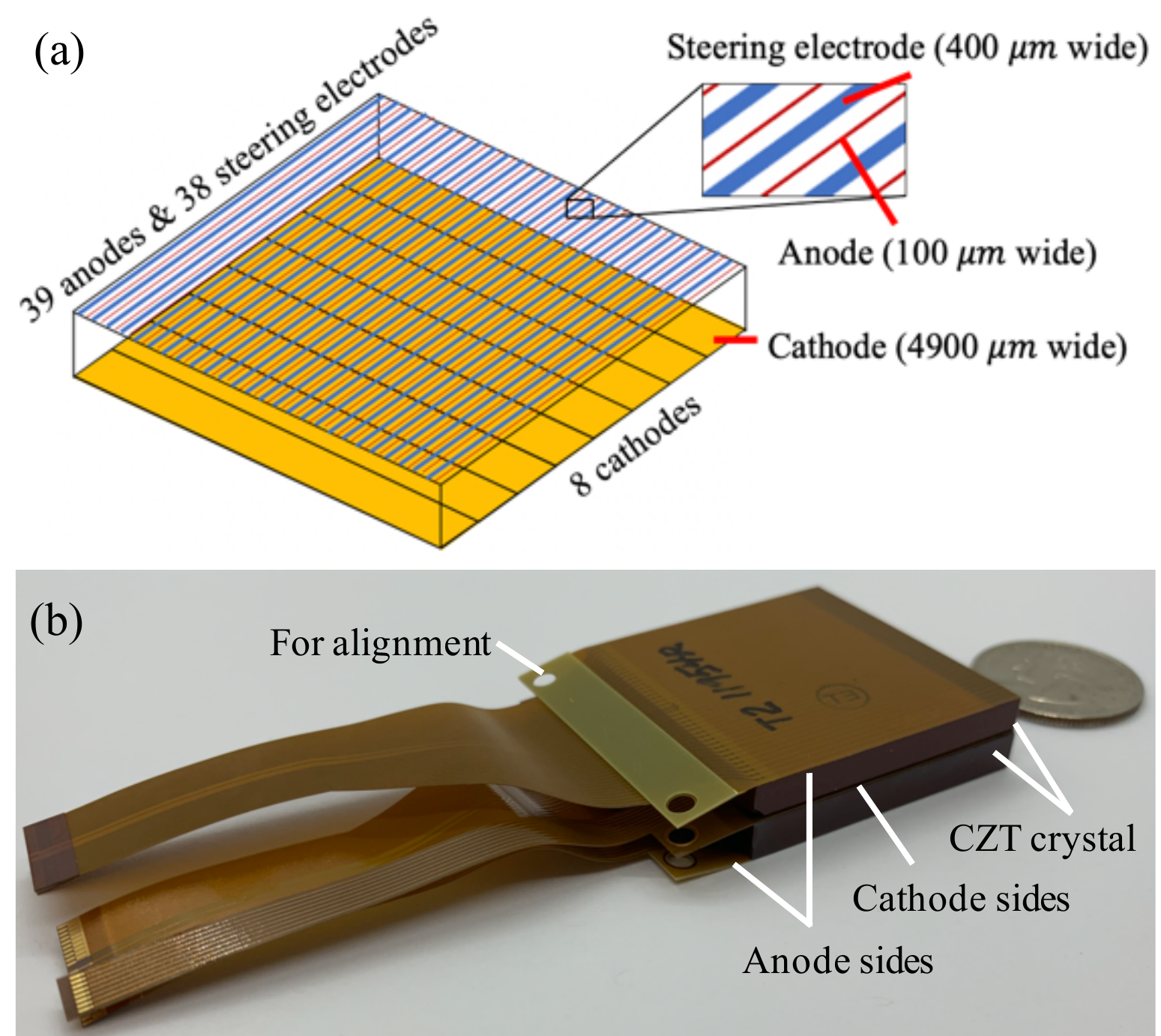}
\end{center}
\caption{(a) Schematic of CZT crystal with cross-strip electrode pattern showing anodes, cathode, and steering electrodes. (b) Two CZT are assembled into a flexible circuit and stacked based on the anode-cathode-cathode-anode configuration to form a CZT module (4 cm $\times$ 4 cm $\times$ 1 cm).}
\label{czt_detector}
\end{figure}

Fig. \ref{czt_detector} (a) shows the design of the CZT detector. Each detector is a monolithic CZT crystal with dimensions of 40 mm $\times$ 40 mm $\times$ 5 mm. Orthogonal 8 cathode electrodes (4900 $\mu$m width) and 39 anode electrodes (100 $\mu$m width) are deposited to the two opposite 40 mm $\times$ 40 mm crystal faces. 38 steering electrodes are interspersed with anodes with the same pitch and with a large width (400 $\mu$m width). During the experiments, a bias of -500V and -80V with respect to anodes is applied to the cathodes and steering electrodes, respectively. The cross-strip electrode pattern was chosen to use fewer electronic readout channels (2n versus n\textsuperscript{2}) \cite{maehlum2007study} while still providing high spatial resolution. The steering electrode was designed for enhancing the anode charge collection.

Two CZT crystals are assembled together using the flexible circuits based on an anode-cathode-cathode-anode (ACCA) stacking structure to form a 40 mm $\times$ 40 mm $\times$ 10 mm CZT module (shown in Fig.\ref{czt_detector} (b)). Conductive silver epoxy is used as the material to facilitate the electrical connections between the CZT crystal and the flexible circuit. This ACCA stacking structure could decrease the dead space between CZT crystals and increase the packing fraction. 

\subsection{Modular readout electronics system}

\begin{figure}[ht]
\begin{center}
\includegraphics[scale=0.65]{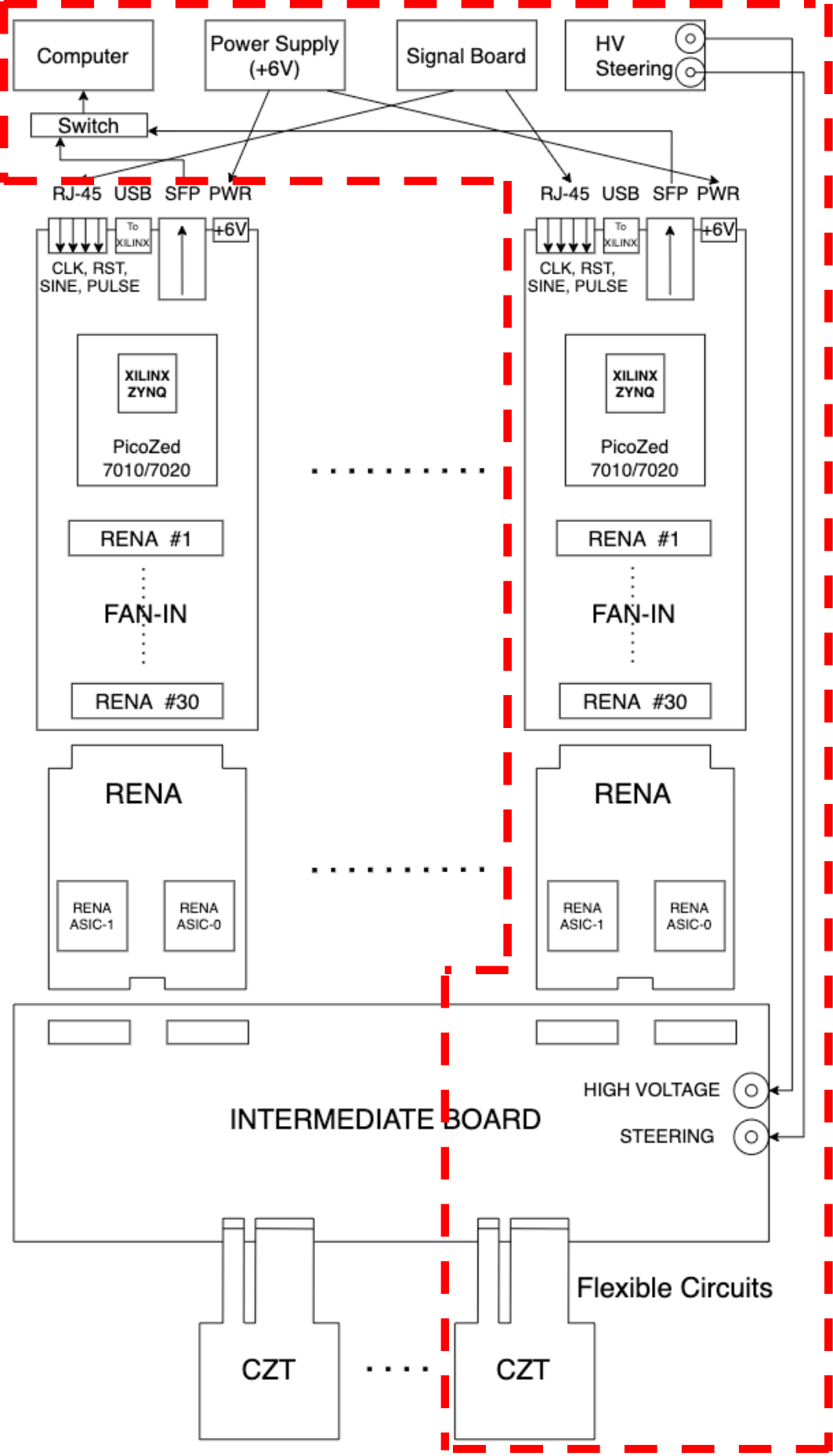}
\end{center}
\caption{Architecture of the DAQ chain of one panel for the head and neck PET system. The red dashed box marked electronic system works as one modular DAQ electronics. Five modular DAQ electronics form the DAQ chain of one panel.}
\label{eletronics_modular}
\end{figure}

\begin{figure}[ht]
\begin{center}
\includegraphics[scale=0.45]{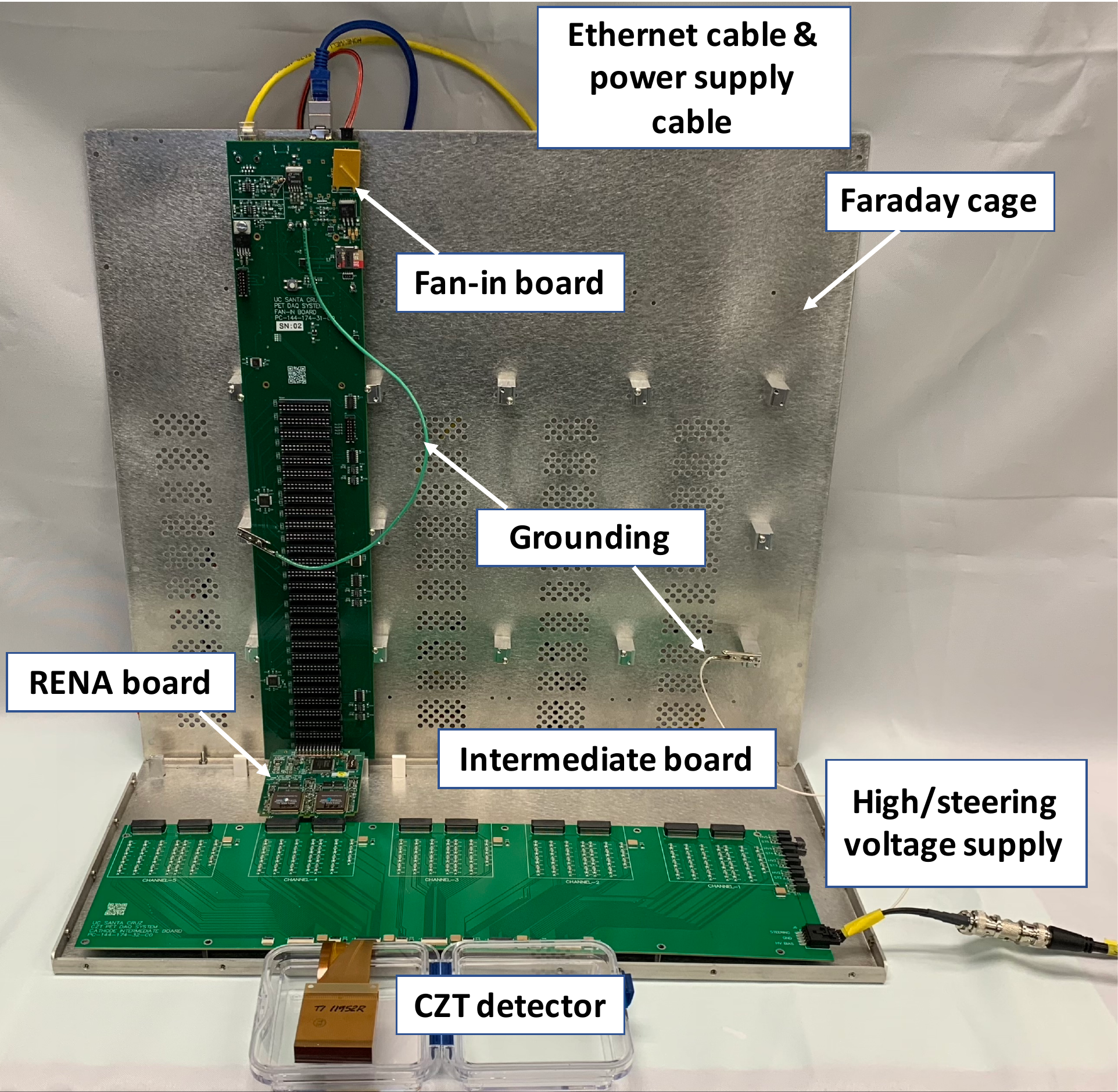}
\end{center}
\caption{An example picture of the experimental setup.}
\label{experiment_setup}
\end{figure}

The architecture schematic of the readout electronic system for one panel is shown in Fig. \ref{eletronics_modular}. The readout system comprised primarily the front-end signal readout part (including the intermediate board and RENA board) and the back-end signal readout part (including the fan-in board and PicoZed board). For investigating the two-channel readout configuration on the cross-strips cadmium zinc telluride detector, the measurement is completed by the red dashed electronic readout system. The experimental setup is shown in Fig. \ref{experiment_setup}.

Each CZT detector has 47 output channels in total (39 anodes and 8 cathodes), which are read by two RENA-3 (Readout Electronics
for Nuclear Applications developed by NOVA R\&D Inc., Riverside, CA) application integrated circuits (ASIC). The RENA-3 ASIC is implemented in a custom-designed front-end board called RENA board. The intermediate board provides the connection to high voltage and steering voltages and also works as the "bridge" to define different readout configurations (half-half or alternative configuration which is presented in detail in \ref{readout_config}) between the CZT detector and the RENA-3 ASIC, which is the focused point of this paper. 

The fan-in board connects to a 1-column by a 30-row array of RENA boards. The fan-in board is responsible for generating the system-wide 50 MHz clock, receiving RENA-3 signals, and distributing the clock and UV signal to all 30 RENA boards. A data acquisition (DAQ) chain, capable of 2 Gbps data transmission, is connected to a DAQ computer via an SFP connector. The DAQ chain was implemented by the Picozed board, which is attached to the backside of the fan-in board.

\subsection{Two intermediate boards with different readout configuration} \label{readout_config}

\begin{figure}[ht]
\begin{center}
\includegraphics[width=\columnwidth]{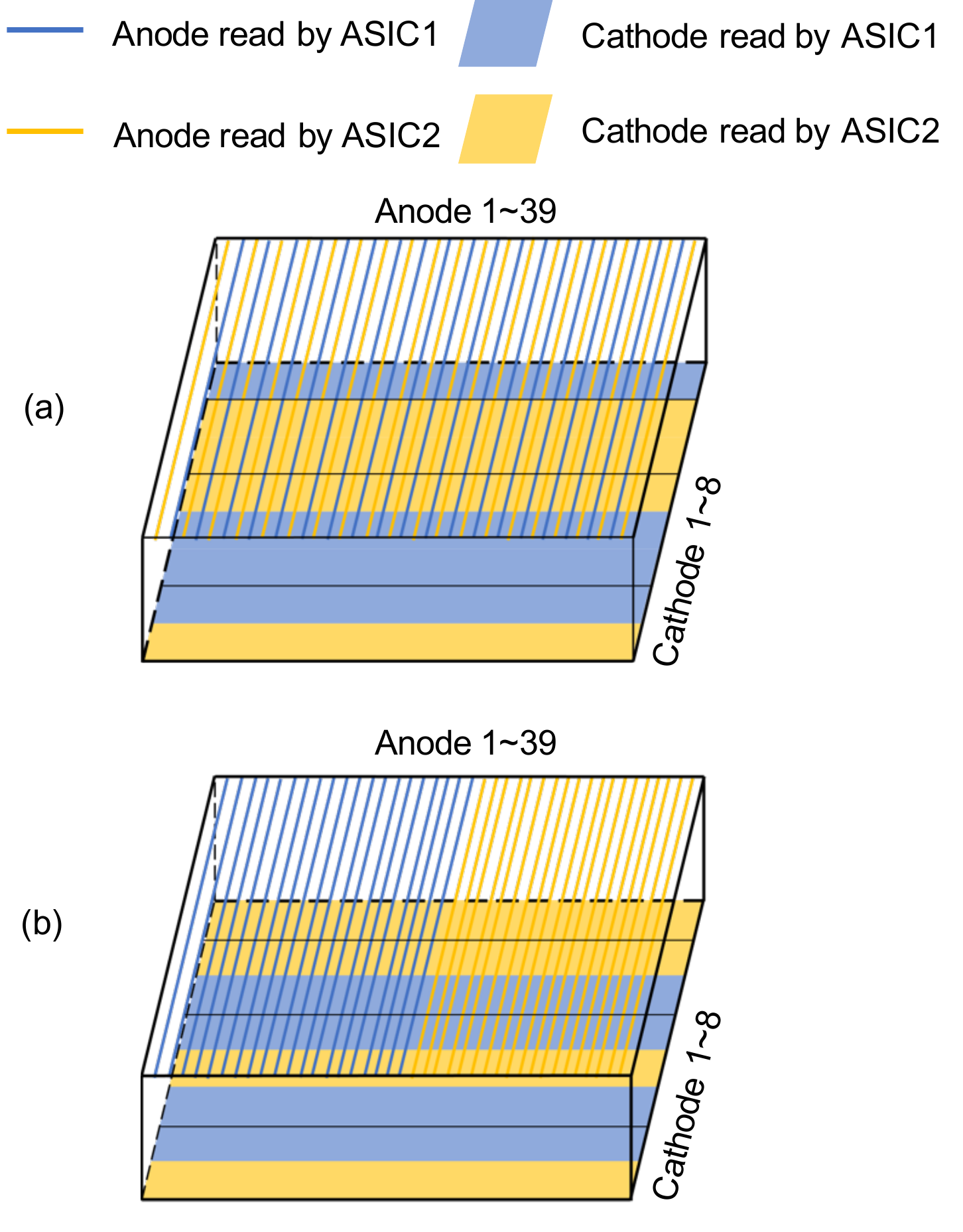}
\end{center}
\caption{Schematic of one CZT electrode distribution to two RENA-3 ASICs readout (a) previous design and for (b) new design in this paper. (a) For the previous design, anodes have the alternative configuration between two RENA-3 ASICs. (b) For the new design, anodes have the half-half configuration where anode 1 to anode 20 are read out with RENA-3 ASIC 1 and anode 21 to anode 39 are readout with RENA-3 ASIC 2.}
\label{readout}
\end{figure}

Two RENA-3 ASICs (36 channels individually) are used to read 39 anodes and 8 cathodes of each CZT detector. Two different readout configurations are applied to anodes: half-half configuration and alternative configuration, which is shown in Fig. \ref{readout}. Alternative readout configuration was the same used in \cite{gu2011study} and \cite{abbaszadeh2016characterization}, where anode 1 was read by RENA-3 ASIC 1 and anode 2 was readout by RENA-3 ASIC 2, and so on. In half-half configuration, anode 1 to anode 20 was read by RENA-3 ASIC 1 (half of the CZT crystal), and RENA-3 ASIC 2 readout anode 21 to anode 39. For the cathode readout configuration, due to the edge-on incident photons, cathodes are with the optimized alternative configuration to share the load between the two RENA-3, which is studied in \cite{li2019influence} by the GATE simulation.

\begin{figure*}[ht]
\begin{center}
\includegraphics[scale=0.5]{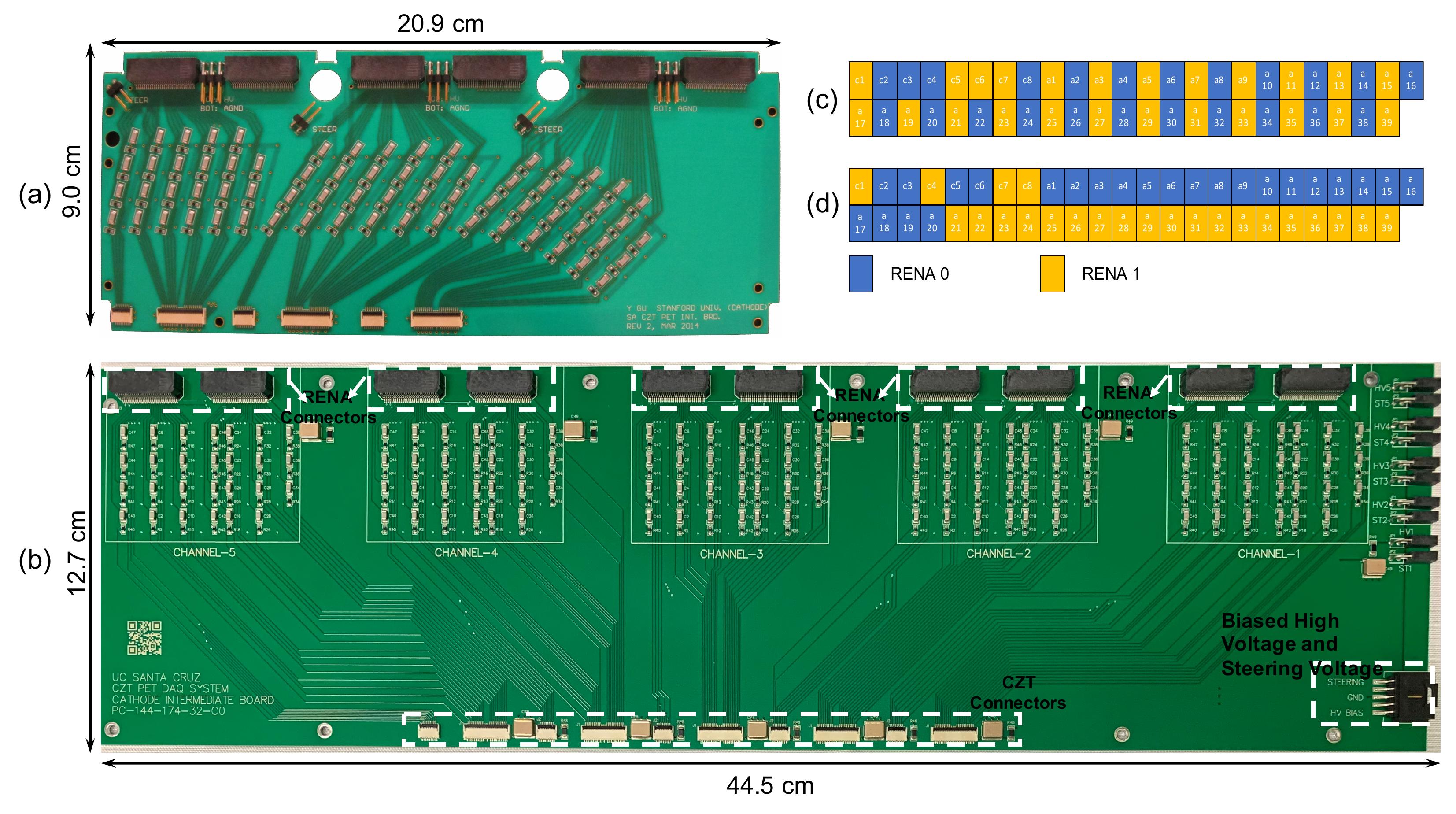}
\end{center}
\caption{(a) Schematics of the intermediate board with alternate readout configuration with dimensions of 9.0 cm $\times$ 20.9 cm. (b) Schematics of the intermediate board with half-half readout configuration with dimensions of 12.7 cm $\times$ 44.5 cm. (c) Channel distribution map between RENA ASICs and CZT anode/cathode readout channels for alternate readout configuration. (d) Channel distribution map between RENA ASICs and CZT anode/cathode readout channels for half-half readout configuration.}
\label{interme}
\end{figure*}

In order to achieve the two different readout configurations, two intermediate boards were designed, shown in Fig. \ref{interme}. Fig. \ref{interme} (a) shows the intermediate board with alternate readout configuration, which is with the size of 9.0 cm $\times$ 20.9 cm to connect one array of 1$\times$3 CZT crystals to three RENA boards. The intermediate boards with half-half configurations are presented in Fig. \ref{interme} (b) to connect 1$\times$5 CZT crystals to five RENA boards. The same CZT connectors and RENA connectors are applied to the two intermediate boards, thus we investigate two aforementioned readout configurations by replacing two intermediate boards.

In order to have a successful operation of the DAQ chain and minimize the time jitters and cross-talk among ASICs channels, the data acquisition trigger thresholds (V$_{thresh}$) for the digital to analog converter (DAC) of each channel on each RENA ASIC was optimized manually. We increased the V$_{thresh}$ value of each channel to above the noise level and thus prevent the signal triggering by the noise. After completing the adjustment of triggering levels for all ASICs, the system is ready for the measurements. The V$_{thresh}$ value of each channel would be monitored and tuned repeatedly when we adjust the readout electronics of the system.

\subsection{Electronics noise with test pulse}\label{noise-test-pulse}

To quantify the internal electronic noise contribution to the energy resolution of the two different intermediate boards (i.e. two different readout configurations), we first used a square wave as a test pulse to provide charge injection to each channel. To simulate equivalent charge injection by a 511 keV photon in a CZT detector, a square wave with 1 kHz with 250 mV peak-to-peak amplitude without offset was used during the experiment. An example of the experimental setup is shown in Fig. \ref{experiment_setup}, which also presents how different boards are connected together. In our studies, data were acquired when:

\begin{itemize}
  \item CZT detector, intermediate board (with half-half readout configuration or with alternative readout configuration), RENA board, and fan-in board are connected together.
  \item HV bias and steering bias of CZT are turned on.
\end{itemize}

Each data acquisition time was set as 5 minutes. The experiment was repeated 5 times to get the standard deviation. The same experimental process is applied to the two intermediate boards with different readout configurations. During the experiments, the whole system is packaged into a light-tight Faraday cage to reduce interference from outside light and external electronic noise. The results of a spectral peak with full width at half maximum (FWHM) were reported in keV units. The corresponding results are shown in section \ref{test-pulse}.

\subsection{Experiments with Ge-68 as the point source}

The anode energy resolution and anode channels' count rate are studied based on the edge-on irradiation configuration (as described here) with a Ge-68 point source. A 50 $\mu$Ci 250 $\mu$m diameter Ge-68 point source was placed 10 mm away from the center of the 40 mm $\times$ 5mm edge-face of the CZT detector. Therefore, the photons from the point source enter the CZT detector by the edge-on configuration and encounter at least 40 mm CZT material, which could yield a detection efficiency greater than 86\% \cite{habte2007effects}. The anode strips of the CZT detector were oriented toward to the point source and the CZT detector was assembled to the head-and-neck system as described in Fig.\ref{eletronics_modular}. 

The data acquisition time of each experiment was set as 10 minutes. The experiment was repeated 5 times to get the standard deviation. The same experimental process is applied to the two intermediate boards with different readout configurations.

The calibration of signal amplitude from the ADC unit to the keV unit was performed for each anode channel. By recording the photopeaks in the ADC unit corresponding to the energy peak at known energies (Ge-68 for 511 keV and Cs-137 for 662 keV), a linear ADC-to-keV calibration map and the ADC/keV conversion factors of each channel were developed. All results reported in the paper were in keV units. 

\subsubsection{Anode energy resolution experiments}

The anode energy resolution refers to an energy spectrum containing “multiple interaction" events deposited energy to each anode channel, which was acquired from the aforementioned experiment process. The “multiple interaction" event contains the photoelectric events (511 keV energy deposition happens to each anode corresponding CZT detection volume), Compton scatters (a large number of interaction events will have energy deposition much less than 511 keV, which corresponds to small amplitude signals and is the dominating events noted in CZT), as well as charge-shared events (interaction happens along the charge detection volume boundary between anodes electrodes). 

Compton scatter is prevalent in CZT detectors, specially designed low-noise, high-sensitivity readout electronics are preferred which enables the signal triggering at lower energy deposition. Moreover, the energy resolution with “multiple interactions" is of particular interest which could allow the system to have higher photon detection sensitivity \cite{abbaszadeh2018positioning}. Thus, we should compare the anode energy resolution performance of the two different readout configurations. The results are summarized in section \ref{anode-energy-resolution}.

In addition, the DAC value of each channel indicates the noise level and thus the lowest detectable energy ability of each channel. The lower the DAC value, the higher the detection sensitivity of the system (i.e. the lower the debatable energy of the “multiple interaction" event). Therefore, we also investigate the lowest DAC value of each readout configuration by lowering the DAC values to just above the noise level. The results are shown in section \ref{anode-energy-resolution}.

\subsubsection{Anode count rate investigations} \label{count_rate}

Readout electronics with high counts rate allow the recording of more valid events per unit of time, which further improve the detection sensitivity. In our study, the CZT detector is with readout channel number larger than that of the RENA ASCIs and thus multiple ASICs are required. For a system with such a readout scheme, load balance is another consideration to achieve a higher count rate. Therefore, we compare the count rate performance of these two readout configurations. Since the interaction with deposited energy between 450 keV to 600 keV is of interest event in PET applications, we only focus on the event in the aforementioned energy threshold. A python code was developed to filter the interested events from the completed events of an anode energy spectrum, which is shown in Fig. \ref{energy_spetra} (d).

To quantitatively analyze the count rate of two different readout configurations, we calculated the count rate value of each anode channel and the total count rate value of all anode channels for each readout configuration. We also calculate the count rate value of each anode channel and the total count rate value of all anode channels for a half-half readout configuration with lower DAC values.

\section{Results} \label{results}

\subsection{Electronics noise with test pulse} \label{test-pulse}

\begin{figure}[ht]
\begin{center}
\includegraphics[width=\columnwidth]{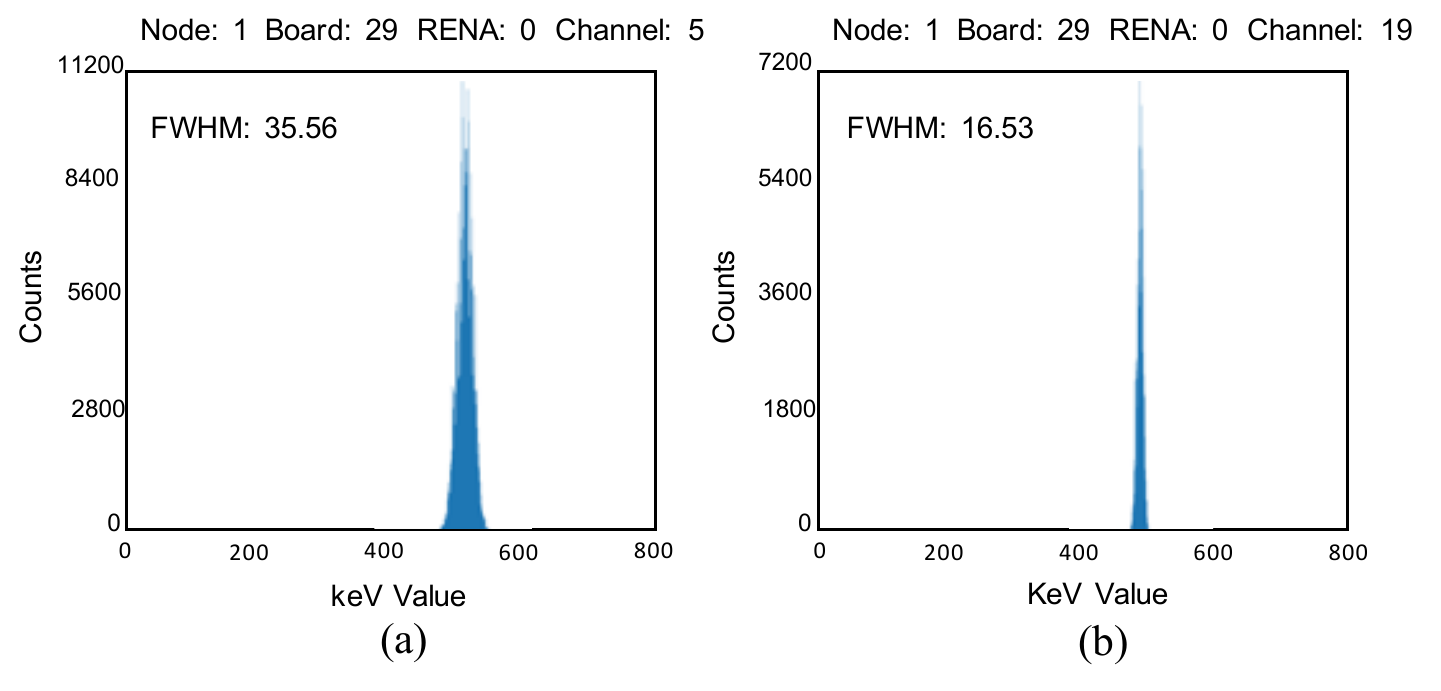}
\end{center}
\caption{Results examples of energy resolution with test pulse for alternate readout configuration in (a) cathode channel and (b) anode channel.}
\label{alternate_test_examples}
\end{figure}

\begin{figure}[ht]
\begin{center}
\includegraphics[width=\columnwidth]{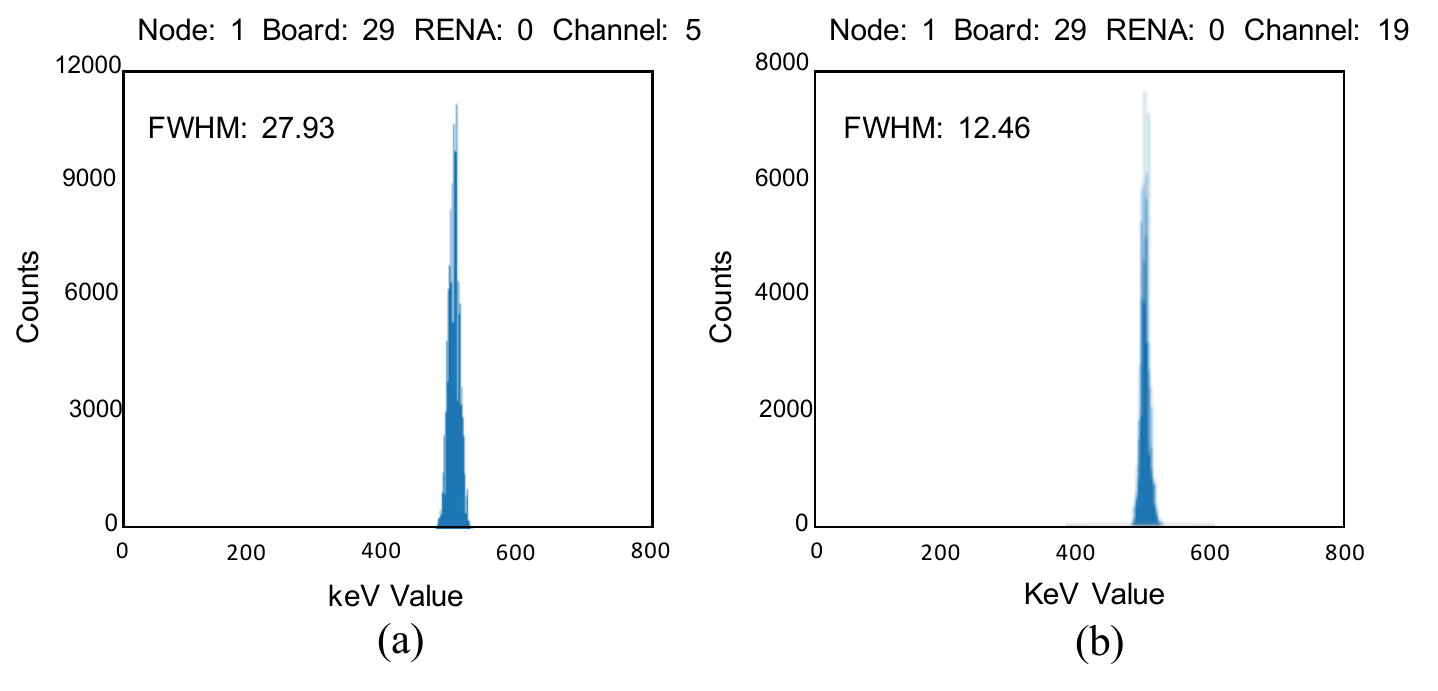}
\end{center}
\caption{Results examples of energy resolution with test pulse for half-half readout configuration in (a) cathode channel and (b) anode channel}
\label{half_test_examples}
\end{figure}

\begin{figure}[ht]
\begin{center}
\includegraphics[width=\columnwidth]{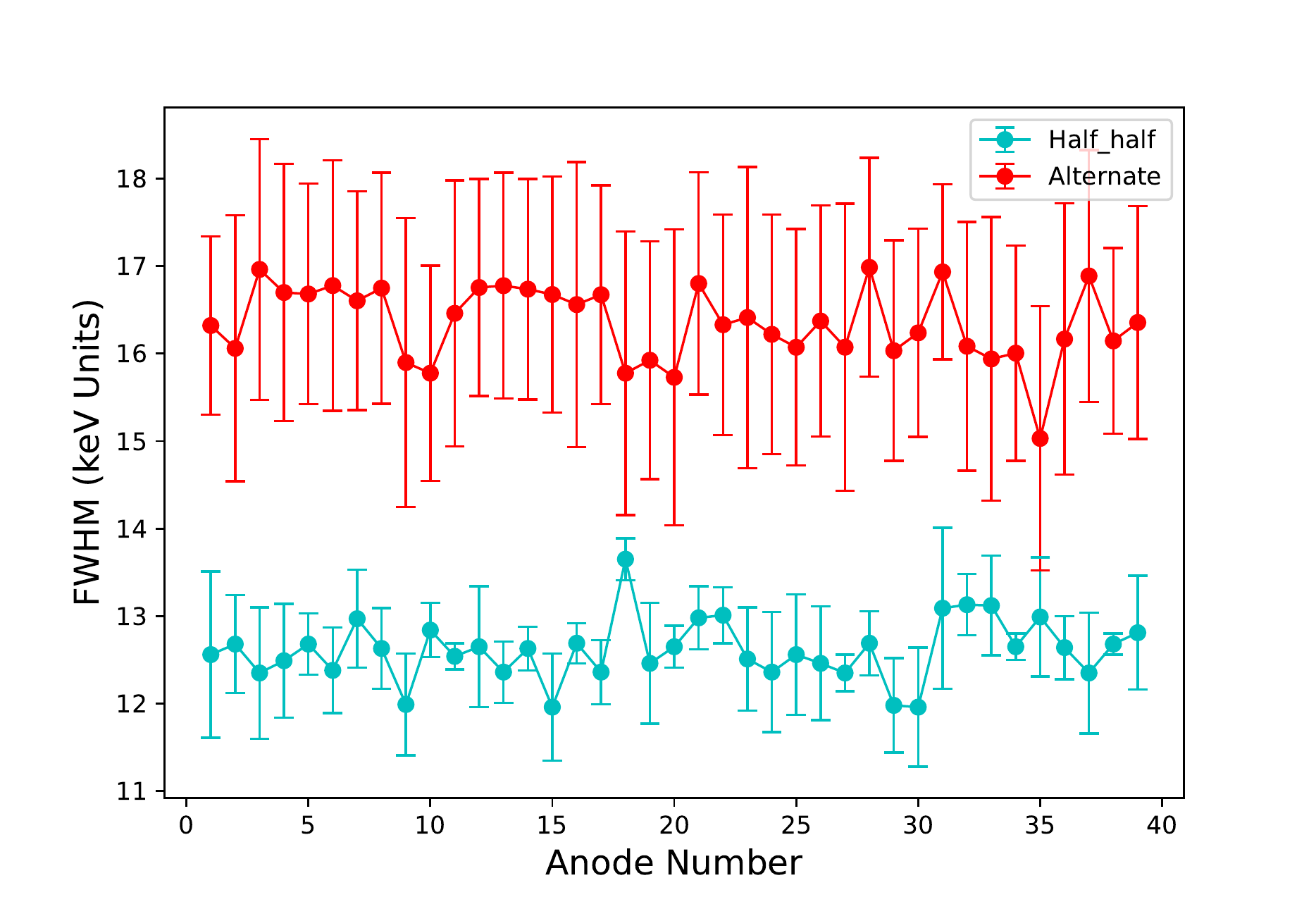}
\end{center}
\caption{Energy resolution comparison for 30 anode channels between alternate readout configuration and half-half readout configuration.}
\label{test_anodes}
\end{figure}

\begin{figure}[ht]
\begin{center}
\includegraphics[width=\columnwidth]{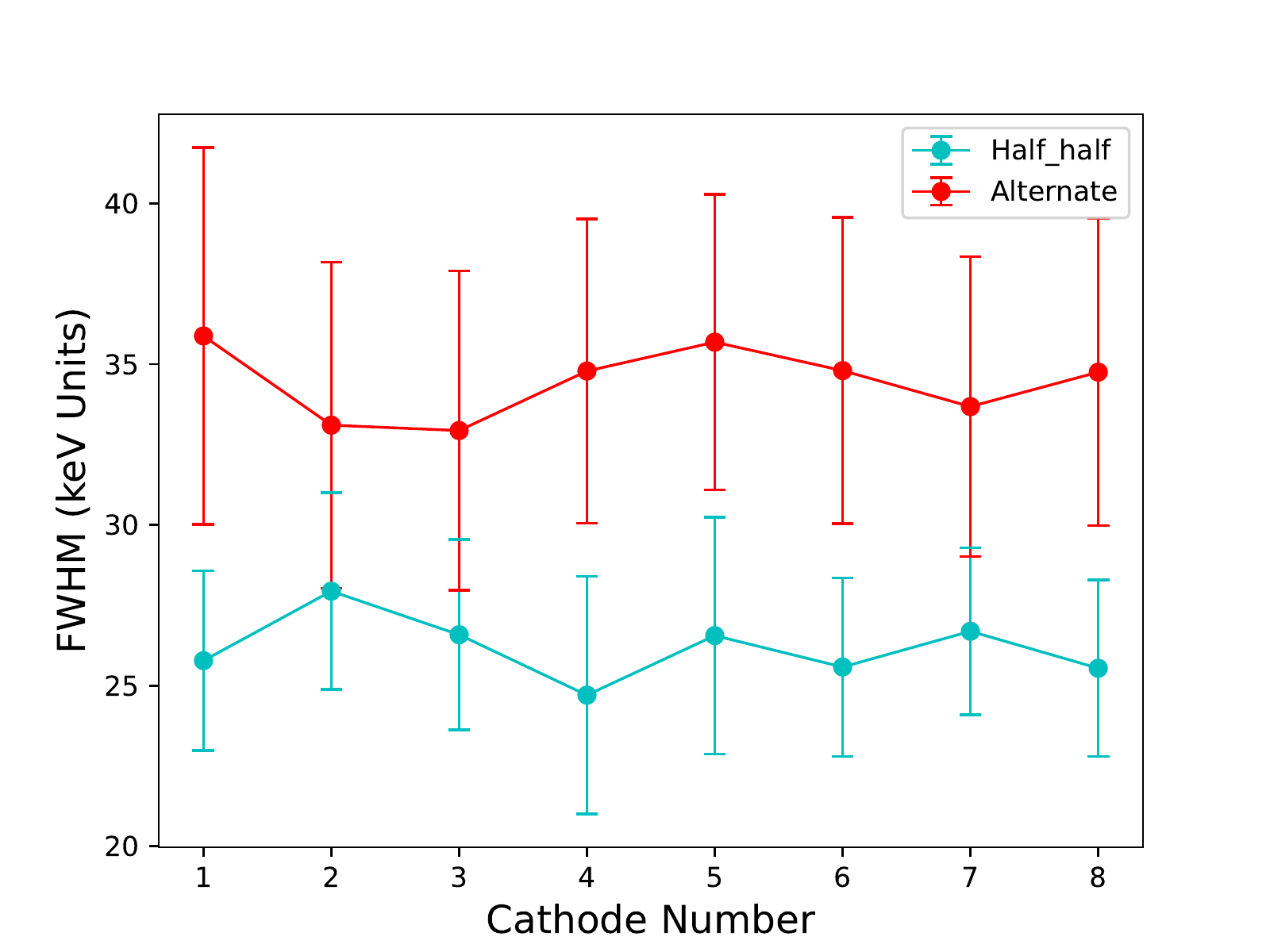}
\end{center}
\caption{Energy resolution comparison for cathode channels between alternate readout configuration and half-half readout configuration.}
\label{test_cathode}
\end{figure}

The result examples of the electronics noise with test pulse using alternate or half-half readout configuration discussed in section \ref{noise-test-pulse} are shown in Fig. \ref{alternate_test_examples} and Fig. \ref{half_test_examples}, respectively. The summary of energy spectra peak with FWHM in keV units for all 39 anode channels with test pulse is shown in Fig. \ref{test_anodes} Fig. \ref{test_cathode} presents the energy spectra peak with FWHM in keV units summary for all 8 cathode channels with test pulse. 

Half-half readout configuration produces the average spectra peak with FWHM of 12.61 $\pm$ 0.48 keV units (anode) units and 26.16 $\pm$ 3.03 keV units (cathode). The average width of the spectra peak broadens to 16.34 $\pm$ 1.37 keV units (anode) and 34.45 $\pm$ 4.92 keV units for alternate readout configuration. 
Overall, for every readout configuration, anode channels are with lower electronic noise level (smaller average FWHM value of the spectra) than cathode channels. The anode/cathode channels from the half-half readout configuration are with lower electronic noise level (smaller average FWHM value of spectra)  than the corresponding anode/cathode channels of alternate readout configuration, which is discussed in section \ref{discussion} in detail. 

\subsection{Energy resolution}\label{anode-energy-resolution}

\begin{figure}[ht]
\begin{center}
\includegraphics[width=\columnwidth]{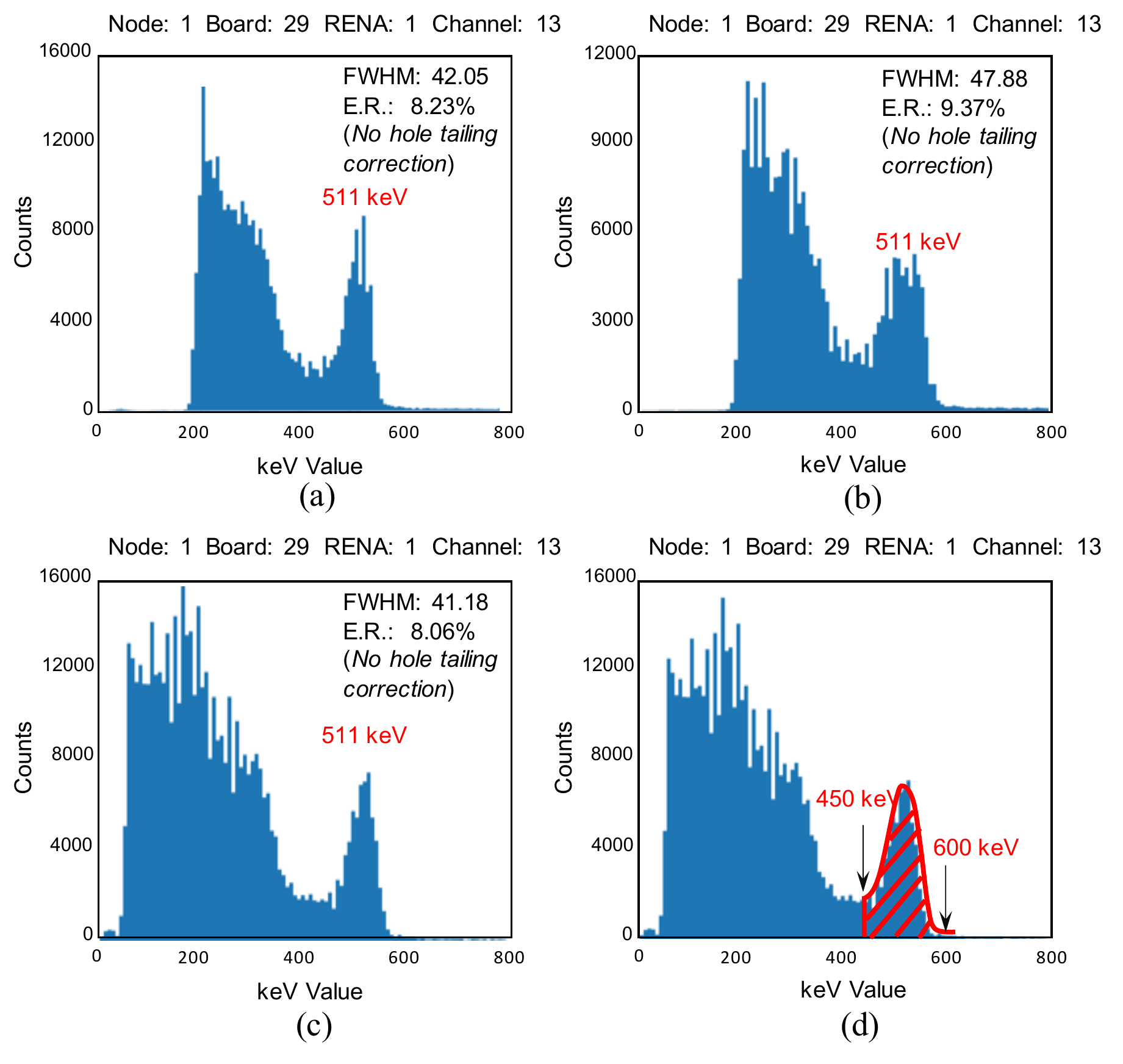}
\end{center}
\caption{(a) Example of one channel energy spectra plot using Ge-68 point source with half-half readout configuration. (b) Example of one channel energy spectra plot using Ge-68 point source with alternate readout configuration. (c) Example of one channel energy spectra plot using Ge-68 point source with half-half readout configuration and lower DAC threshold. (d) Examples showing the count rate between 450 keV and 600 keV.}
\label{energy_spetra}
\end{figure}

Fig. \ref{energy_spetra} (a) and (b) show the examples of energy spectra for alternate and half-half readout configurations using Ge-68 as the point source. The DAC values of Fig. \ref{energy_spetra} (a) and (b) are set as 65 ADC units, which are the lowest achievable DAC values of alternate readout configuration. The energy spectra with lower DAC values (55 ADC units, the lowest achievable DAC values for the half-half readout configurations) using Ge-68 for the half-half readout configuration are presented in Fig. \ref{energy_spetra} (c). The lowest detectable energy ability of alternate readout configuration is around 200 keV units, while that of half-half readout configuration is lower than 50 keV units.

\begin{figure}[ht]
\begin{center}
\includegraphics[width=\columnwidth]{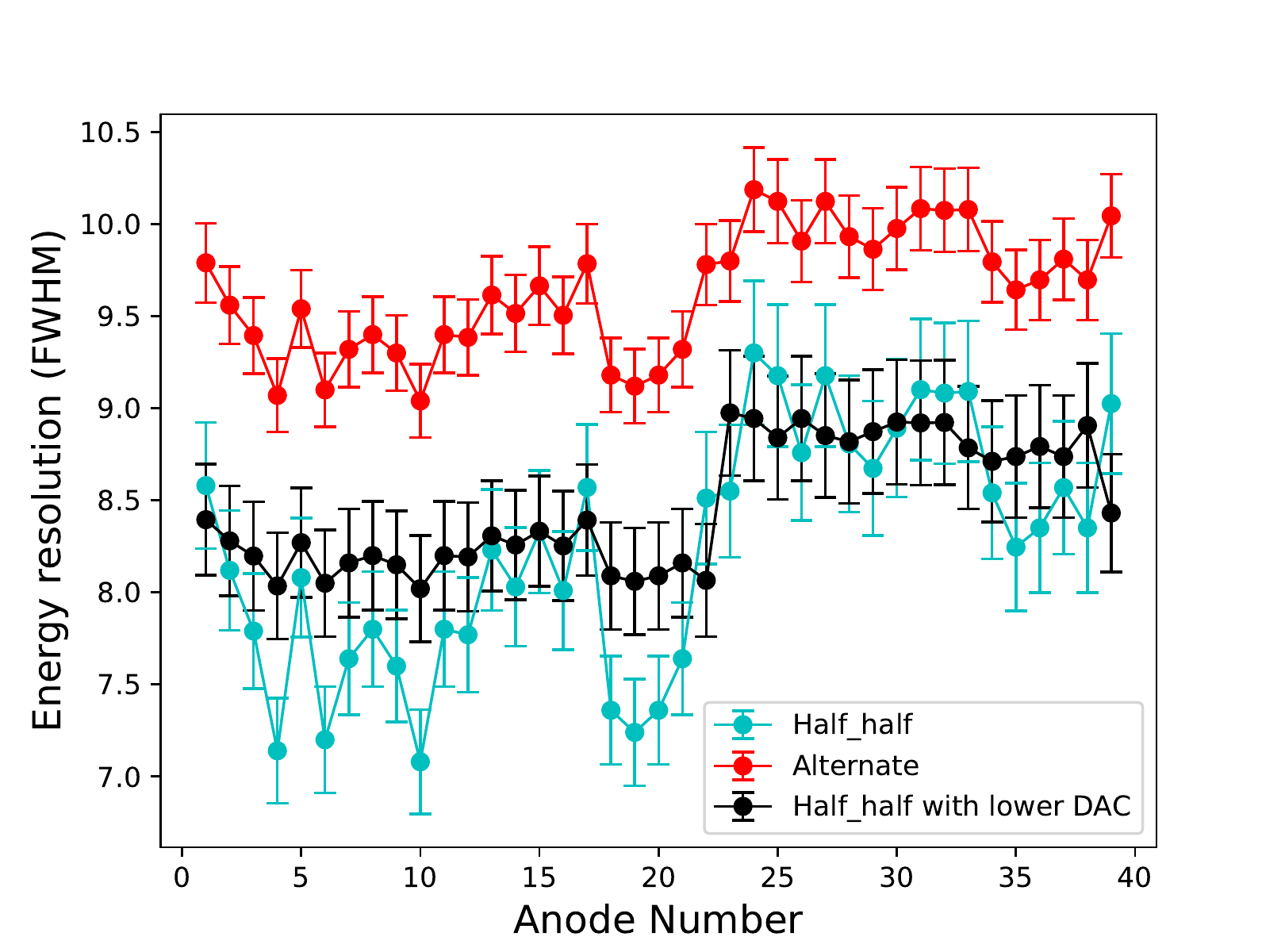}
\end{center}
\caption{Summary of energy resolution for half-half readout configuration, alternate readout configuration, and half-half readout configuration with lower DAC threshold.}
\label{energy_resolution_summary}
\end{figure}

The anode channels' FWHM energy resolution summary of the above-mentioned three experiments scenarios is reported in Fig. \ref{energy_resolution_summary}. All presented results are without tailing correction. From Fig. \ref{energy_resolution_summary}, we could know that the FWHM energy resolution is 9.72 $\pm$ 0.33 in keV units (1.90\% $\pm$ 0.06\%) for alternate readout configuration, 8.45 $\pm$ 0.21 in keV units (1.65\% $\pm$ 0.05\%) for half-half readout configuration and 8.68 $\pm$ 0.31 in keV units (1.69\% $\pm$ 0.06\%) for half-half readout configuration with lower DAC values. It's noticeable that half-half readout configurations (with or without lower DAC values) have better performance on energy resolution than the alternate readout configuration. Moreover, The energy resolution performance of center anode channels (anode channel No. 18 to No. 22) for all three tested readout conditions is better than the sides anode channels, which is further discussed in section \ref{discussion}.

\subsection{Count rate}

\begin{figure}[ht]
\begin{center}
\includegraphics[width=\columnwidth]{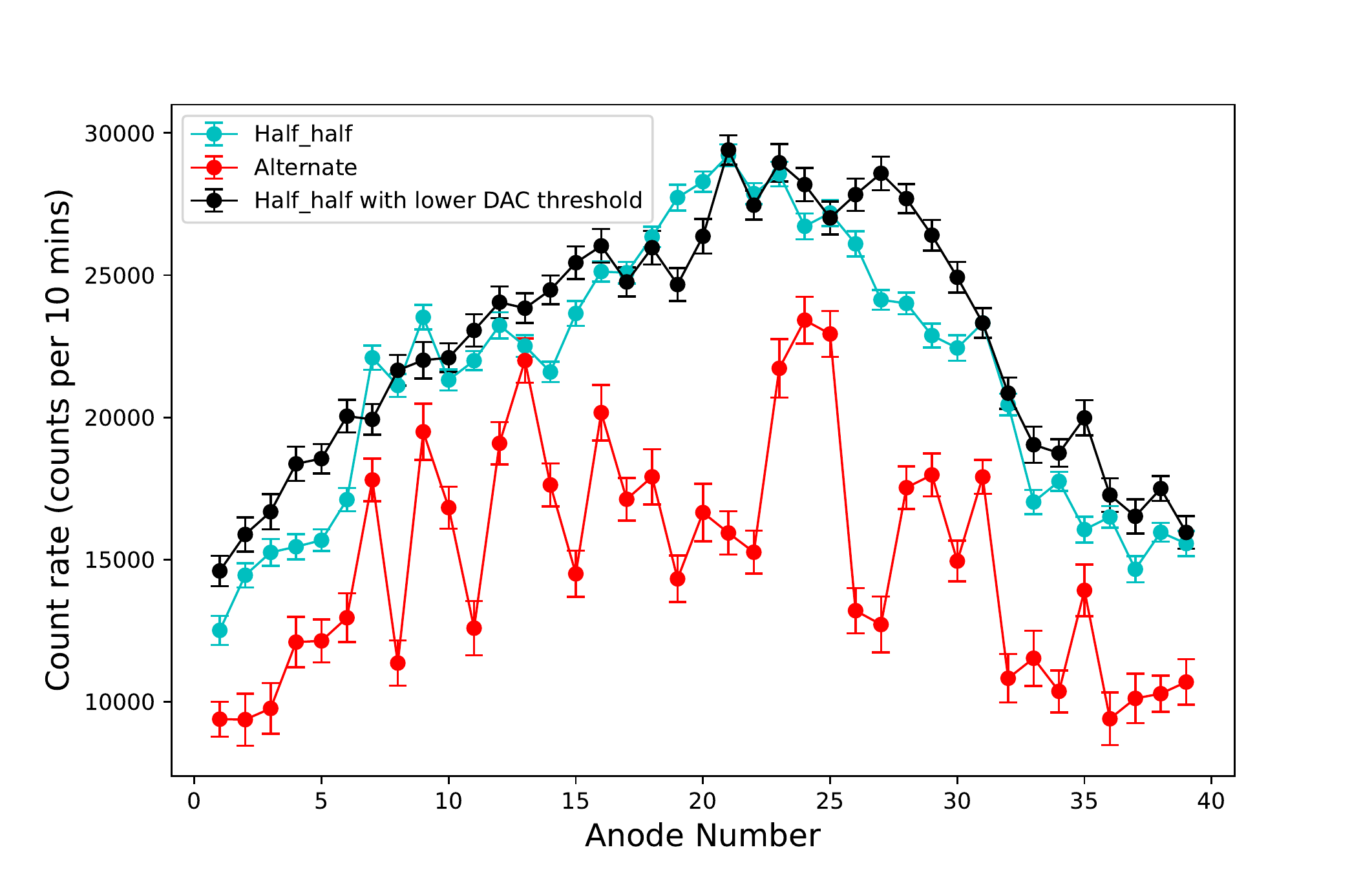}
\end{center}
\caption{Summary of count rate for half-half readout configuration, alternate readout configuration and half-half readout configuration with lower DAC threshold.}
\label{Count_3}
\end{figure}

According to the method described in section \ref{count_rate}, we acquired Fig. \ref{Count_3}, which shows the count rate for alternate, half-half, and half-half with lower DAC threshold readout configurations. From Fig. \ref{Count_3}, it can be seen that half-half or the half-half with lower DAC values readout configurations are with comparable count rate results, which are markedly higher than that of alternate readout configurations. This point will be discussed in section \ref{discussion}. The average count's rate values for all 39 anode channels are 14972.56 $\pm$ 826.29 (alternate), 21550.58 $\pm$ 409.94 (half-half), 22669.62 $\pm$ 561.63 (half-half with lower DAC).

The comparison between the previous simulation work \cite{li2019influence} and the experimental studies in this paper will also be discussed in detail in section \ref{discussion}.

\section{Discussion} \label{discussion}

\subsection{Analysis on electronics noise}


From the internal comparison of Fig. \ref{alternate_test_examples} (a) and (b) or Fig. \ref{half_test_examples} (a) and (b), we could see a lower electronics noise from the anode electrode than that from the cathode electrode. The width of cathode electrode strips is 8 times the width of anode electrode strips. For the collection of the mobile holes, larger cathode electrodes are preferable to maintain high photon sensitivity, while at the cost of introducing higher leakage or dark current of the CZT detector. The cathode electrode also has higher sensitivity to random fluctuation in the current  amplitude due to the electronic noise. Therefore, a larger electronic noise is expected in the cathode electrode.


All presented data (Fig. \ref{test_anodes} Fig. \ref{test_cathode}) were acquired with exactly the same experimental conditions, except for the different readout configurations and the dimension of intermediate boards. For the factor of the intermediate board, since the intermediate board with a half-half readout configuration is with a larger dimension than that board with an alternate readout configuration, a higher electronic noise level should be expected in the intermediate board with a half-half readout configuration. However, from Fig. \ref{test_anodes} and Fig. \ref{test_cathode}, we could see a lower electronics noise level (small FWHM value) in alternate readout configuration and higher electronics noise level in half-half readout configuration, for both anode electrodes and cathode electrodes. Therefore, we could conclude a much lower electronics noise level of the half-half readout configuration than that of the alternate readout configuration.

\subsection{Analysis on energy resolution}


As shown in Fig. \ref{energy_spetra}, DAC values setting in half-half readout configuration (65 ADC units in Fig. \ref{energy_spetra} (c)) is lower than that of alternate readout configuration (55 ADC units in Fig. \ref{energy_spetra} (a)). DAC value refers to the trigger threshold level of each ASIC's digital-to-analog converter. The DAC values can be found by the empirical mean and should be set just above the electronic noise level, specifically, the valid trigger data could be taken at the chosen trigger threshold level while all electronic noise could be avoided. Therefore, the noise level of each channel is the decisive factor of its DAC value. From the section \ref{test-pulse}, the half-half readout configuration with a lower noise level, thus a lower DAC value.


From Fig. \ref{energy_resolution_summary}, a better energy resolution performance is shown in the half-half readout configuration when with the same DAC values as the alternate readout configuration, which is mainly due to a low electronics noise level of half-half readout configuration when compared to the alternate readout configuration. By comparing the results of half-half/half-half with lower DAC in Fig. \ref{energy_resolution_summary}, we could see a smaller variance in the result of half-half readout configuration with lower DAC. ASIC readout channels maintain a higher detection sensitivity and higher count rate with a lower DAC value setting, therefore, the variance of the results could be relatively reduced. From Fig. \ref{energy_resolution_summary}, we also could observe that the best energy resolution performance for each readout configuration is always from No. 18 to No. 22 anode channels. The location of the point isotope source, which is placed in the detector center, corresponding to the anode channels around No. 20, contributes to this phenomenon. 



\subsection{Analysis on count rate}


Two observations can be made from Fig. \ref{Count_3}. Firstly, it is clear that the count rate, in the half-half configuration, is higher than the that of alternate configuration. Half-half configurations are with close count rates with different DAC values. Secondly, the alternate configuration showed a serrated shape and a large variance between neighboring channels. The count rate performance in the half-half configuration is more consistent. These two aforementioned observations echos the results of our previous simulation paper \cite{li2019influence}. 

Compared to the count rate of the alternate configuration, the count rate of the half-half configuration improved by 43.9\% and 51.4\% (with low DAC values), respectively. Charge sharing and scattering are two main factors contributing to the different count rates. In the CZT detector, due to the carrier density gradient and electrostatic repulsion, the induced charge cloud will expand during the drift to the collection electrode. The possible largest root means the square radius of the charge cloud can be 100 $\mu$m \cite{benoit2009simulation}. In our designed CZT detector, the distance of neighboring anodes was 1 mm, thus the charge sharing should contribute to the count rate and needs to be considered. Compare to the alternate configuration, the only charge-sharing situation happens to the two anodes in the middle where two neighboring anodes were monitored by two different RENA-3 ASICs. Therefore, the half-half configuration is much less susceptible to charge sharing than the alternate configuration. For the scattering, since the average scattering length of a 511 keV photon was about 7 mm, thus it was difficult for a scattered photon to escape from one half to the other half (20 mm length for each half of the detector). As a result, the alternate configuration was more vulnerable to scattering.

\section{Conclusion}
In this work, the influence of channel configuration between two ASICs on the electronics noise, FWHM energy resolution, and count rate of a CZT detector with a cross-strip pattern was studied. A lower electronics noise and better FWHM energy resolution were observed in the detector with a half-half readout configuration. Due to charge sharing and scattering, the half-half configuration of anode channels showed a higher count rate than the alternate configuration. With the half-half anode configuration, the count was 43.9\% higher than that of the alternate anode configuration.

\bibliographystyle{IEEEtran}
\bibliography{mybib}

\end{document}